\pdfoutput=1
\documentclass[12pt,a4paper]{article}

\usepackage{ifthen} 
\usepackage{dcolumn,booktabs}
\newboolean{pdflatex}
\setboolean{pdflatex}{true} 

\newboolean{articletitles}
\setboolean{articletitles}{true} 

\newboolean{uprightparticles}
\setboolean{uprightparticles}{false} 

\newboolean{inbibliography}
\setboolean{inbibliography}{false} 

\def\paperauthors{LHCb Collaboration} 
\def\papertitle{} 
\def\papercopyright{\the\year\ CERN for the benefit of the LHCb Collaboration} 
\def\paperlicence{CC BY 4.0 licence}
\def\paperlicenceurl{https://creativecommons.org/licenses/by/4.0/}

\def\Lcstar      {{\ensuremath{\Lz^{*+}_\cquark}}\xspace}

\usepackage{microtype}
\usepackage{lineno}  
\usepackage{xspace} 
\usepackage{caption} 

\usepackage{graphicx}  
\usepackage{color}
\usepackage{colortbl}
\graphicspath{{./figs/}} 

\usepackage{amsmath} 
\usepackage{amssymb}
\usepackage{amsfonts}
\usepackage{upgreek} 

\newcommand*\patchAmsMathEnvironmentForLineno[1]{%
\expandafter\let\csname old#1\expandafter\endcsname\csname #1\endcsname
\expandafter\let\csname oldend#1\expandafter\endcsname\csname
end#1\endcsname
 \renewenvironment{#1}%
   {\linenomath\csname old#1\endcsname}%
   {\csname oldend#1\endcsname\endlinenomath}%
}
\newcommand*\patchBothAmsMathEnvironmentsForLineno[1]{%
  \patchAmsMathEnvironmentForLineno{#1}%
  \patchAmsMathEnvironmentForLineno{#1*}%
}
\AtBeginDocument{%
\patchBothAmsMathEnvironmentsForLineno{equation}%
\patchBothAmsMathEnvironmentsForLineno{align}%
\patchBothAmsMathEnvironmentsForLineno{flalign}%
\patchBothAmsMathEnvironmentsForLineno{alignat}%
\patchBothAmsMathEnvironmentsForLineno{gather}%
\patchBothAmsMathEnvironmentsForLineno{multline}%
\patchBothAmsMathEnvironmentsForLineno{eqnarray}%
}

\usepackage{hyperref}    
\usepackage[all]{hypcap} 

\usepackage{xspace} 
\usepackage{upgreek}

\def\lhcb {\mbox{LHCb}\xspace}

\def\babar  {\mbox{BaBar}\xspace}
\def\belle  {\mbox{Belle}\xspace}

\def\MagUp {\mbox{\em Mag\kern -0.05em Up}\xspace}

\ifthenelse{\boolean{uprightparticles}}%
{

 \def\Pmu         {\ensuremath{\upmu}\xspace}                 
 \def\Pnu         {\ensuremath{\upnu}\xspace}                 
                  
 \def\Ppi         {\ensuremath{\uppi}\xspace}                 
                  
 \def\Prho        {\ensuremath{\uprho}\xspace}                 
                  
 \def\Ptau        {\ensuremath{\uptau}\xspace}

 \def\PDelta      {\ensuremath{\Delta}\xspace}                 
 \def\PXi      {\ensuremath{\Xi}\xspace}                 
 \def\PLambda      {\ensuremath{\Lambda}\xspace}                 
 \def\PSigma      {\ensuremath{\Sigma}\xspace}                 
 \def\POmega      {\ensuremath{\Omega}\xspace}                 
 \def\PUpsilon      {\ensuremath{\Upsilon}\xspace}                 
 
 \def\PB      {\ensuremath{\mathrm{B}}\xspace}                 
                  
 \def\PD      {\ensuremath{\mathrm{D}}\xspace}

 \def\PK      {\ensuremath{\mathrm{K}}\xspace}

 \def\Pb      {\ensuremath{\mathrm{b}}\xspace}                 
 \def\Pc      {\ensuremath{\mathrm{c}}\xspace}

 \def\Pi      {\ensuremath{\mathrm{i}}\xspace}

 \def\Ps      {\ensuremath{\mathrm{s}}\xspace}

}
{

 \def\Pmu         {\ensuremath{\mu}\xspace}                 
 \def\Pnu         {\ensuremath{\nu}\xspace}                 
                  
 \def\Ppi         {\ensuremath{\pi}\xspace}                 
                  
 \def\Prho        {\ensuremath{\rho}\xspace}                 
                  
 \def\Ptau        {\ensuremath{\tau}\xspace}

 \mathchardef\PDelta="7101
 \mathchardef\PXi="7104
 \mathchardef\PLambda="7103
 \mathchardef\PSigma="7106
 \mathchardef\POmega="710A
 \mathchardef\PUpsilon="7107
                  
 \def\PB      {\ensuremath{B}\xspace}                 
                  
 \def\PD      {\ensuremath{D}\xspace}

 \def\PK      {\ensuremath{K}\xspace}

 \def\Pb      {\ensuremath{b}\xspace}                 
 \def\Pc      {\ensuremath{c}\xspace}

 \def\Pi      {\ensuremath{i}\xspace}

 \def\Ps      {\ensuremath{s}\xspace}

}

\makeatletter
\ifcase \@ptsize \relax
  \newcommand{\miniscule}{\@setfontsize\miniscule{4}{5}}
\or
  \newcommand{\miniscule}{\@setfontsize\miniscule{5}{6}}
\or
  \newcommand{\miniscule}{\@setfontsize\miniscule{5}{6}}
\fi
\makeatother

\DeclareRobustCommand{\optbar}[1]{\shortstack{{\miniscule (\rule[.5ex]{1.25em}{.18mm})}
  \\ [-.7ex] $#1$}}

\def\mun        {{\ensuremath{\Pmu^-}}\xspace} 

\def\tauon      {{\ensuremath{\Ptau}}\xspace}

\def\taum       {{\ensuremath{\Ptau^-}}\xspace}

\def\neu        {{\ensuremath{\Pnu}}\xspace}
\def\neub       {{\ensuremath{\overline{\Pnu}}}\xspace}

\def\neumb      {{\ensuremath{\neub_\mu}}\xspace}
\def\neut       {{\ensuremath{\neu_\tau}}\xspace}
\def\neutb      {{\ensuremath{\neub_\tau}}\xspace}

\def\squark    {{\ensuremath{\Ps}}\xspace}

\def\cquark    {{\ensuremath{\Pc}}\xspace}

\def\bquark    {{\ensuremath{\Pb}}\xspace}

\def\pion   {{\ensuremath{\Ppi}}\xspace}
\def\piz    {{\ensuremath{\pion^0}}\xspace}

\def\pip    {{\ensuremath{\pion^+}}\xspace}
\def\pim    {{\ensuremath{\pion^-}}\xspace}

\def\rhomeson {{\ensuremath{\Prho}}\xspace}
\def\rhoz     {{\ensuremath{\rhomeson^0}}\xspace}

\def\kaon    {{\ensuremath{\PK}}\xspace}
  \def\Kbar    {{\kern 0.2em\overline{\kern -0.2em \PK}{}}\xspace}

\def\KorKbar    {\kern 0.18em\optbar{\kern -0.18em K}{}\xspace}

\def\Kp      {{\ensuremath{\kaon^+}}\xspace}
\def\Km      {{\ensuremath{\kaon^-}}\xspace}

  \def\Dbar    {{\kern 0.2em\overline{\kern -0.2em \PD}{}}\xspace}
\def\D       {{\ensuremath{\PD}}\xspace}

\def\DorDbar    {\kern 0.18em\optbar{\kern -0.18em D}{}\xspace}

\def\Dzb     {{\ensuremath{\Dbar{}^0}}\xspace}

\def\Dm      {{\ensuremath{\D^-}}\xspace}

\def\Dstarp  {{\ensuremath{\D^{*+}}}\xspace}

\def\Ds      {{\ensuremath{\D^+_\squark}}\xspace}

\def\Dsm     {{\ensuremath{\D^-_\squark}}\xspace}

\def\Dssm    {{\ensuremath{\D^{*-}_\squark}}\xspace}

\def\B       {{\ensuremath{\PB}}\xspace}
\def\Bbar    {{\ensuremath{\kern 0.18em\overline{\kern -0.18em \PB}{}}}\xspace}
\def\Bb      {{\ensuremath{\Bbar}}\xspace}
\def\BorBbar    {\kern 0.18em\optbar{\kern -0.18em B}{}\xspace}

\def\Bzb     {{\ensuremath{\Bbar{}^0}}\xspace}

  \def\Y#1S{\ensuremath{\PUpsilon{(#1S)}}\xspace}

\def\Lz          {{\ensuremath{\PLambda}}\xspace}
\def\Lbar        {{\ensuremath{\kern 0.1em\overline{\kern -0.1em\PLambda}}}\xspace}
\def\LorLbar    {\kern 0.18em\optbar{\kern -0.18em \PLambda}{}\xspace}

\def\Lb      {{\ensuremath{\Lz^0_\bquark}}\xspace}

\def\Lc      {{\ensuremath{\Lz^+_\cquark}}\xspace}

\newcommand{\decay}[2]{\ensuremath{#1\!\to #2}\xspace}         

\def\to                 {\ensuremath{\rightarrow}\xspace}

\def\qsq       {{\ensuremath{q^2}}\xspace}

\def\AT#1     {\ensuremath{A_{\mathrm{T}}^{#1}}\xspace}           

\def\C#1      {\ensuremath{\mathcal{C}_{#1}}\xspace}                       
\def\Cp#1     {\ensuremath{\mathcal{C}_{#1}^{'}}\xspace}                    
\def\Ceff#1   {\ensuremath{\mathcal{C}_{#1}^{\mathrm{(eff)}}}\xspace}        
\def\Cpeff#1  {\ensuremath{\mathcal{C}_{#1}^{'\mathrm{(eff)}}}\xspace}       
\def\Ope#1    {\ensuremath{\mathcal{O}_{#1}}\xspace}                       
\def\Opep#1   {\ensuremath{\mathcal{O}_{#1}^{'}}\xspace}                    

\newcommand{\tev}{\ifthenelse{\boolean{inbibliography}}{\ensuremath{~T\kern -0.05em eV}}{\ensuremath{\mathrm{\,Te\kern -0.1em V}}}\xspace}
\newcommand{\gev}{\ensuremath{\mathrm{\,Ge\kern -0.1em V}}\xspace}
\newcommand{\mev}{\ensuremath{\mathrm{\,Me\kern -0.1em V}}\xspace}
\newcommand{\kev}{\ensuremath{\mathrm{\,ke\kern -0.1em V}}\xspace}
\newcommand{\ev}{\ensuremath{\mathrm{\,e\kern -0.1em V}}\xspace}
\newcommand{\gevc}{\ensuremath{{\mathrm{\,Ge\kern -0.1em V\!/}c}}\xspace}
\newcommand{\mevc}{\ensuremath{{\mathrm{\,Me\kern -0.1em V\!/}c}}\xspace}
\newcommand{\gevcc}{\ensuremath{{\mathrm{\,Ge\kern -0.1em V\!/}c^2}}\xspace}
\newcommand{\gevgevcccc}{\ensuremath{{\mathrm{\,Ge\kern -0.1em V^2\!/}c^4}}\xspace}
\newcommand{\mevcc}{\ensuremath{{\mathrm{\,Me\kern -0.1em V\!/}c^2}}\xspace}

\def\mum  {\ensuremath{{\,\upmu\mathrm{m}}}\xspace}

\def\invfb   {\ensuremath{\mbox{\,fb}^{-1}}\xspace}

\newcommand{\chisq}{\ensuremath{\chi^2}\xspace}

\def\gsim{{~\raise.15em\hbox{$>$}\kern-.85em
          \lower.35em\hbox{$\sim$}~}\xspace}
\def\lsim{{~\raise.15em\hbox{$<$}\kern-.85em
          \lower.35em\hbox{$\sim$}~}\xspace}

\def\sqs   {\ensuremath{\protect\sqrt{s}}\xspace}

\def\pt         {\mbox{$p_{\mathrm{ T}}$}\xspace}

\def\evtgen     {\mbox{\textsc{EvtGen}}\xspace}

\def\geant      {\mbox{\textsc{Geant4}}\xspace}

\def\photos     {\mbox{\textsc{Photos}}\xspace}

\def\pythia     {\mbox{\textsc{Pythia}}\xspace}

\def\tell1  {TELL1\xspace}
\def\ukl1   {UKL1\xspace}

\newcommand{\ie}{\mbox{\itshape i.e.}\xspace}

\usepackage{cite} 
\usepackage{mciteplus}

\usepackage{float}

\usepackage{longtable} 

\begin{document}

\renewcommand{\thefootnote}{\fnsymbol{footnote}}
\setcounter{footnote}{1}


\renewcommand{\thefootnote}{\arabic{footnote}}
\setcounter{footnote}{0}

\cleardoublepage


\pagestyle{plain} 
\setcounter{page}{1}
\pagenumbering{arabic}




\def\papertitle{
Observation of the decay $\Lb\to\Lc\taum\neutb$}

\renewcommand{\thefootnote}{\fnsymbol{footnote}}
\setcounter{footnote}{1}


\begin{titlepage}
\pagenumbering{roman}

\vspace*{-1.5cm}
\centerline{\large EUROPEAN ORGANIZATION FOR NUCLEAR RESEARCH (CERN)}
\vspace*{1.5cm}
\noindent
\begin{tabular*}{\linewidth}{lc@{\extracolsep{\fill}}r@{\extracolsep{0pt}}}
\ifthenelse{\boolean{pdflatex}}
{\vspace*{-1.5cm}\mbox{\!\!\!\includegraphics[width=.14\textwidth]{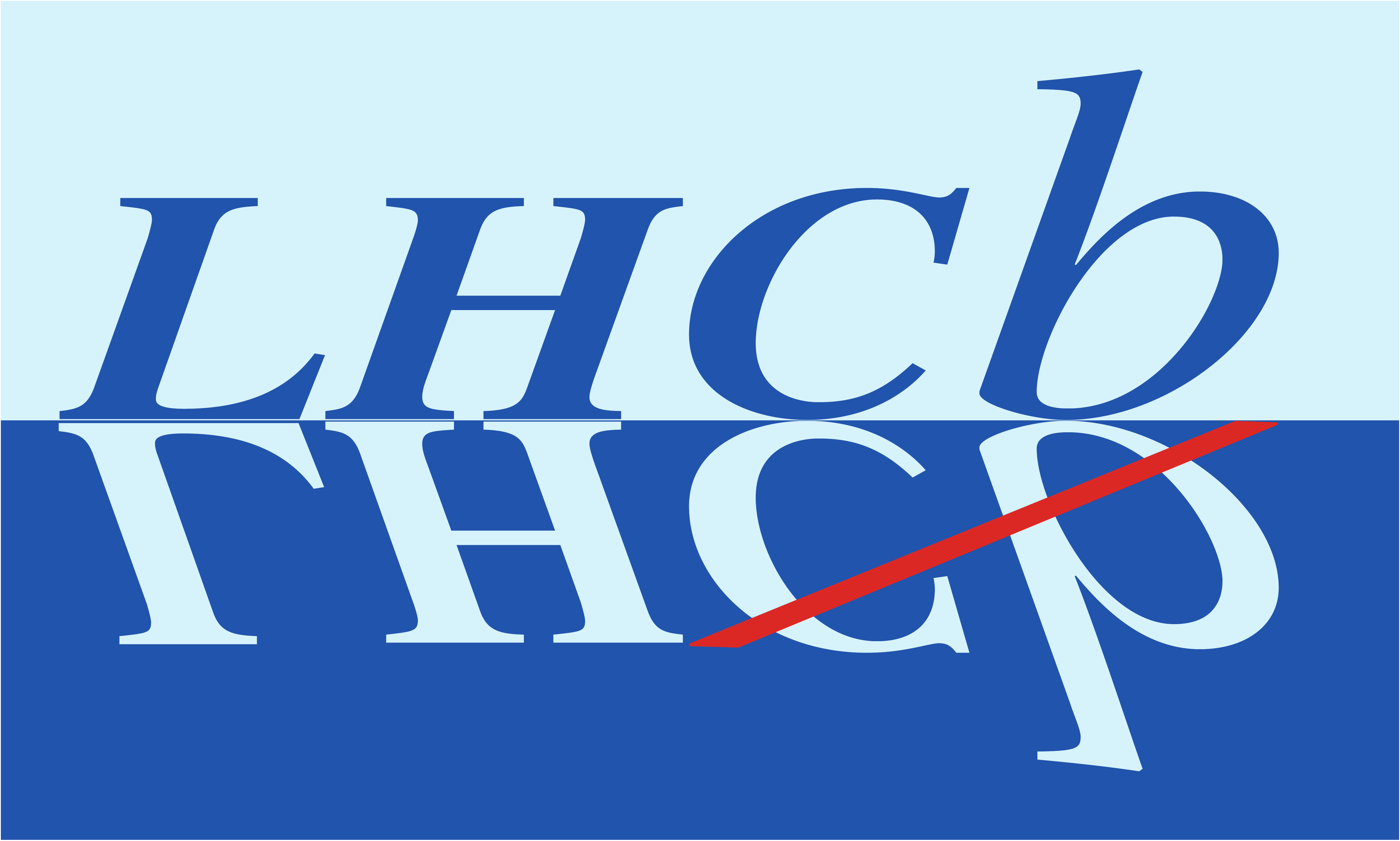}} & &}%
{\vspace*{-1.2cm}\mbox{\!\!\!\includegraphics[width=.12\textwidth]{figs/lhcb-logo.eps}} & &}%
\\
 & & CERN-EP-2021-265 \\  
 & & LHCb-PAPER-2021-044 \\  
 & & May 13, 2022\\ 
 & & \\
\end{tabular*}

\vspace*{2.0cm}

{\normalfont\bfseries\boldmath\huge
\begin{center}
  \papertitle 
\end{center}
}

\vspace*{2.0cm}

\begin{center}
\paperauthors\footnote{Authors are listed at the end of this Letter.}
\end{center}

\vspace{\fill}

\begin{abstract}
  \noindent
 The first observation of the semileptonic $b$-baryon decay $\Lb \to \Lc \tau^-\overline{\nu}_{\tau}$, with a significance of $6.1\,\sigma$, is reported using a data
    sample corresponding to 3\invfb  of integrated luminosity, collected by
    the LHCb experiment at centre-of-mass energies of 7 and 8~TeV at the LHC. The \taum lepton is reconstructed in the hadronic
    decay to three charged pions. The ratio ${\cal{K}}={\cal{B}}(\Lb \to \Lc \taum
    \overline{\nu}_{\tau})/{\cal{B}}(\Lb \to \Lc \pi^-\pi^+\pi^-)$ is measured to be  ${\ensuremath 2.46 \pm 0.27\pm 0.40}$, where the first uncertainty is statistical and the second systematic. The branching fraction  ${\cal{B}}(\Lb \to \Lc\tau^-\overline{\nu}_{\tau}) = {\ensuremath (1.50 \pm 0.16\pm 0.25\pm 0.23)\%}$ is obtained, where the third uncertainty is from the external branching fraction of the normalization channel \mbox{\Lb\to\Lc\pim\pip\pim}. The ratio   of semileptonic branching fractions ${\cal{R}}(\Lc)\equiv {\cal{B}}(\Lb \to \Lc \taum
    \overline{\nu}_{\tau})/{\cal{B}}(\Lb \to \Lc \mu^-\overline{\nu}_{\mu})$ is derived to be ${\ensuremath  0.242 \pm 0.026 \pm 0.040\pm 0.059}$, where the external branching fraction uncertainty from the channel \Lb\to\Lc\mun\neumb contributes to the last term. This result is  in  agreement with the standard model prediction. 
\end{abstract}

\vspace*{2.0cm}

\begin{center}
  Published in Phys. Rev. Lett. 128, 191803.
\end{center}

\vspace{\fill}

{\footnotesize 
\centerline{\copyright~\papercopyright. \href{\paperlicenceurl}{\paperlicence}.}}
\vspace*{2mm}

\end{titlepage}


\newpage
\setcounter{page}{2}
\mbox{~}
%
%
%
%


\renewcommand{\thefootnote}{\arabic{footnote}}
\setcounter{footnote}{0}

\cleardoublepage


\pagestyle{plain} 
\setcounter{page}{1}
\pagenumbering{arabic}

In the standard model of particle physics (SM) 
flavor-changing processes, such as semileptonic decays of \bquark hadrons, are mediated by  $W^\pm$ 
bosons with universal coupling to leptons. 
Differences in the rates of decays involving the three lepton families  are expected to arise only from
the different masses of the charged leptons. Lepton flavor universality  can be violated in
many extensions of the SM with non-standard flavor
structure. 
Since the uncertainty due to hadronic effects cancels to a large
extent, the SM predictions for the ratios between branching fractions of semileptonic
decays of $b$ hadrons, such as {\mbox{${\cal{R}}(D^{(*)})\equiv {\cal{B}}(\Bb \to D^{(*)} \taum\neutb)/{\cal{B}}(\Bb \to D^{(*)}\mun\neumb)$}}~\cite{Bigi:2016mdz,Jaiswal:2020wer,Bernlochner2017}, where  $D^{(*)}$ and $B$ mesons can be either charged or neutral,
and  {\mbox{${\cal{R}}(\Lambda_{c}^{+})\equiv {\cal{B}}(\Lb \to \Lc \taum
    \overline{\nu}_{\tau})/{\cal{B}}(\Lb \to \Lc \mu^-\overline{\nu}_{\mu})$}}, 
are known with uncertainties at the per cent
level ~\cite{Berna, Bernlochner, Datta}.  These ratios therefore provide a sensitive probe of SM extensions ~\cite{Datta, rmp}.

Measurements of
${\cal{R}}(D^{0,+})$ and ${\cal{R}}(D^{*+,0})$ 
with \taum decay  final states involving electrons or muons have been reported by the
\babar~\cite{Lees:2012xj,Lees:2013uzd} and 
\belle~\cite{Huschle:2015rga,Belle:2019rba,bellepol} Collaborations.  The \lhcb
Collaboration published a determination of
${\cal{R}}(D^{*+})$~\cite{LHCb-PAPER-2015-025},  where the $\tau$ lepton is reconstructed using leptonic
decays to a muon. The LHCb experiment has also reported a measurement of ${\cal{R}}(\Dstarp)$
using the three-prong decay $\taum\to\pim\pip\pim(\piz)\nu_{\tau}$ 
\cite{LHCb-PAPER-2017-027}. 
These ${\cal{R}}(D^{(*)+,0})$ measurements yield values that are larger than the
SM predictions with a combined significance of
3.4 standard deviations  ($\sigma$) to date ~\cite{HFLAV18}. 

This Letter reports the observation of the decay \Lb\to\Lc\taum$\overline{\nu}_{\tau}$ and the first determination of ${\cal{R}}(\Lc)$
using  $\taum\to\pim\pip\pim (\piz)\nu_{\tau}$ decays. The inclusion of charge-conjugate modes is implied throughout. The present work closely follows the strategy of Ref.~\cite{LHCb-PAPER-2017-027}.
Measurements in the baryonic sector  provide complementary constraints on a potential lepton flavor universality violation because of the half-integer spin of the initial state~\cite{Bernlochner, Datta}. 
The \Lb\to\Lc transition is determined by a different set of form factors with respect to the mesonic decays probed so far. Likewise, new physics  couplings can also be different, 
resulting in different scenarios regarding deviations from SM expectations of $\mathcal{R}$(\Lc) and $\mathcal{R}(D^{(*)}$)~\cite{rmp}.  

A data sample of proton-proton ($pp$)
collisions at centre-of-mass energies \sqs=
7 and 8 TeV, corresponding to an integrated luminosity of 3\invfb, 
collected with the \lhcb detector is used. The \lhcb detector is a 
single-arm forward spectrometer covering the pseudorapidity range 
$2 < \eta < 5$, 
described in detail in Refs.~\cite{Alves:2008zz,LHCb-DP-2014-002}.  The detector includes a high-precision tracking system
consisting of a silicon-strip vertex detector surrounding the $pp$
interaction region~\cite{LHCb-DP-2014-001}, and large-area silicon-strip detectors located upstream and downstream  of the $4{\mathrm{\,Tm}}$ dipole magnet.
The minimum distance of a track to a primary $pp$ collision vertex (PV), the impact parameter (IP), 
is measured with a resolution of $(15+29/\pt)\mum$,
where \pt is the component of the momentum transverse to the beam direction, in\,\gevc. 
The online event selection is performed by a
trigger system~\cite{LHCb-DP-2012-004}, 
which consists of a hardware stage based on information from the calorimeter and muon
systems, followed by a software stage that performs a full event
reconstruction. Events are selected at the hardware stage if the particles forming
the signal candidate satisfy  a requirement on the energy deposited in the calorimeters or
if any other particles pass any trigger algorithm. The software trigger requires a two-, three-, or four-track secondary
vertex with significant displacement from any PV and consistent with the decay of a \bquark hadron, or
a three-track vertex with a significant displacement from any PV
and consistent with the decay of a \Lc baryon. A multivariate
algorithm  is used for the identification of secondary vertices consistent with
the decay of a $b$ hadron, while secondary vertices consistent with the
decay of a \Lc baryon are identified using topological  
criteria.
In the simulation, $pp$ collisions are generated using
\pythia8~\cite{Sjostrand:2006za,*Sjostrand:2007gs} 
with a specific \lhcb
configuration~\cite{LHCb-PROC-2010-056}.  Decays of hadronic particles
are described by \evtgen~\cite{Lange:2001uf}, in which final-state
radiation is generated using \photos~\cite{davidson2015photos}. 
The {\mbox{\textsc{Tauola}}\xspace} package~\cite{Davidson:2010rw} is used
to simulate the decays of the \taum lepton into 3\pion\neut and
3\pion\piz\neut final states, where $3\pion\equiv\pim\pip\pim$, according to the resonance chiral Lagrangian model~\cite{Nugent:2013hxa} with a tuning
based on the results from the \babar Collaboration~\cite{Nugent:2013ij}. 
The interaction of the generated particles with the detector, and its response,
are implemented using the \geant
toolkit~\cite{Allison:2006ve, *Agostinelli:2002hh} as described in
Ref.~\cite{LHCb-PROC-2011-006}. The signal decays are simulated using form factors that are derived
from heavy-quark effective theory~\cite{Caprini:1997mu}.

The \Lc baryon candidates are reconstructed using 
the \decay{\Lc}{pK^-\pip} decay mode, by combining three charged tracks compatible with proton, kaon and pion hypotheses. The  \taum candidates are formed by \pim\pip\pim combinations and include contributions from  the \decay{\taum}{3\pion\nu_{\tau}} and
\decay{\taum}{3\pion\piz\nu_{\tau}} decay modes, as neutral pions are
not reconstructed. The \Lc and \taum candidates are selected based
on kinematic, geometric, and particle-identification criteria. The \Lb candidate is formed by combining  a \Lc and a \taum candidate.  Background due to misreconstructed $b$ hadrons, where at least
one additional particle originates from either the 3\pion vertex or the
$b$-hadron vertex, is suppressed by requiring a single \Lb candidate
per event.  Tracks other than those used for the signal candidate are exploited in  a multivariate algorithm to assess the signal isolation, \ie the absence of extra tracks compatible with the 3\pion vertex~\cite{LHCb-PAPER-2015-031}.
The algorithm is  trained on simulated samples of  \mbox{\Lb\to\Lc\taum\neutb} and \mbox{\Lb\to\Lc\Dzb\Km} decays for signal and background, respectively; its efficiency is 20\% higher  
than the cut-based algorithm used in Ref.~\cite{LHCb-PAPER-2017-027} for the same rejection factor. Likewise, the neutral-particle energy contained in a cone centred around the 
direction of the \taum candidates is used to further separate signal and background processes.
The \taum momentum can be determined, up to a two-fold ambiguity, from the momentum
vector of the 3\pion system and the flight direction of the \taum 
candidate. The average
of the two solutions is used, as discussed in Ref.~\cite{LHCb-PAPER-2017-027}. The same method is used to compute the \Lb momentum. This enables the computation of the invariant mass squared of the $\taum\neutb$ lepton pair (\qsq),  and the pseudo decay time of the \taum candidate ($t_{\tauon}$). The variables \qsq and $t_{\tauon}$ are reconstructed with a resolution of roughly 15\%, providing good discrimination between the signal and background processes.

The finite \taum lifetime causes the 3\pion vertex to be detached from the \Lb vertex. This key feature allows the suppression of the large background, called prompt background hereafter, from $b$ hadrons decaying to a \Lc baryon accompanied by a 3\pion system  being produced promptly at the $b$-hadron decay vertex, plus any other unreconstructed particles $(X)$. The difference of the positions of the 3\pion and the \Lc vertices along the beam direction is required
to be at least 5 times larger than its uncertainty. This requirement suppresses the prompt background 10 times more than the selection used in Ref.\cite{LHCb-PAPER-2017-027}, reducing this initially dominant background  to a negligible level, at a price of 50\% reduction of the signal efficiency.

Double-charm background processes due to \Lb baryon decays into a \Lc baryon plus another charmed hadron that subsequently decays into a final state containing three charged pions, are topologically similar to the signal and  constitute the largest background source. The main contribution originates from \decay{\Lb}{\Lc\Dsm(X)}
decays, with \Dsm decays to $3\pion Y$ final states, where $Y$ stands for any set of extra particles, such as one or two \piz mesons. Such \Dsm decays have a large branching fraction ($\sim$30\%)~\cite{PDG2020}. This background  is reduced principally by taking into account the resonant structure of the 3\pion system. The \mbox{$\taum\to3\pion\neut$} decays proceed predominantly through the
$a_1(1260)^-\to\rhoz\pim$ decay. By contrast, the \Dsm\to\mbox{$3\pion Y$} decays occur mainly through the $\eta$ and $\eta^\prime$ resonances. This feature, captured by the shapes of the distributions of the smaller and larger  mass of the two \pip\pim combinations extracted from each 3\pion candidate, the energy carried by neutral particles within the cone around the 3\pion direction, and kinematic variables from partial reconstruction are exploited by means of a 
boosted decision tree (BDT) classifier ~\cite{Breiman,AdaBoost}, as described in Ref.~\cite{LHCb-PAPER-2017-027}. Fig.~\ref{figsupp:bdt_input} of Supplemental Material~\cite{supplbdt} displays the markedly different distributions of the three main input variables to the BDT classifier obtained for signal and \Dsm background, respectively. The partial reconstruction of the \Lb decay kinematics is performed  under the background hypothesis
where the \Lb particle decays to $\Lc\Dsm (\to 3\pi Y)$. 
The BDT response in simulation is validated using three control samples: the $\Lb\to\Lc 3\pion$ normalization sample; a $\Lb\to\Lc\Dzb (X)$ data sample with the subsequent 
\decay{\Dzb}{K^{+}3\pion} decay, which is obtained by
removing the charged-particle isolation criterion and requiring an
additional charged kaon originating from the 3\pion vertex; and a $\Lb\to\Lc\Dm (X)$ data sample, using \Dm\to\Kp\pim\pim decays, which is 
obtained by assigning   a kaon mass to  the positively-charged pion of the \taum candidate. For all these samples, good
agreement between data and simulation is
observed in the distributions of the variables used in the BDT classifier. 

\begin{figure}[!htbp]
   \begin{center}
 \includegraphics[width=0.49\textwidth]{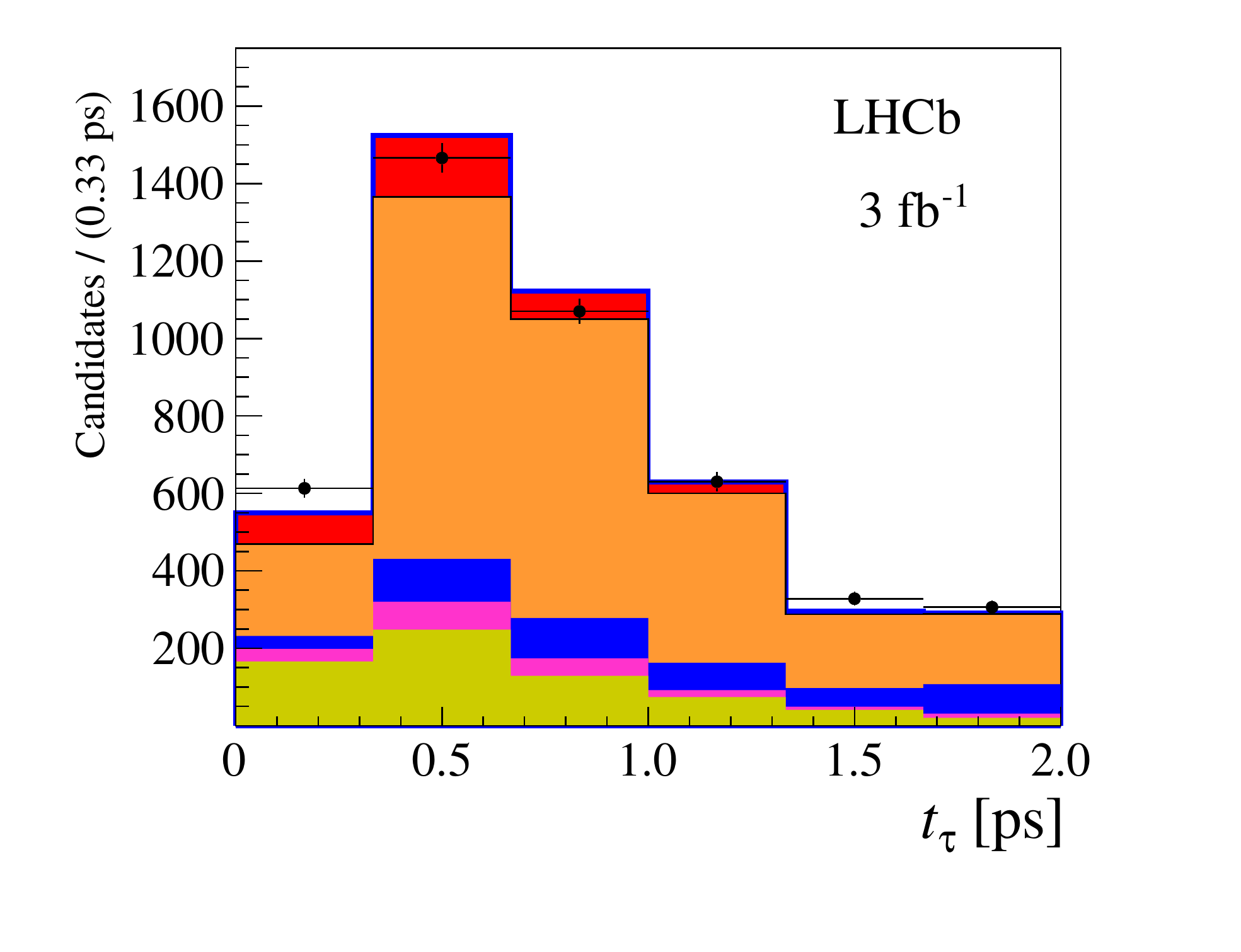}
 \includegraphics[width=0.49\textwidth]{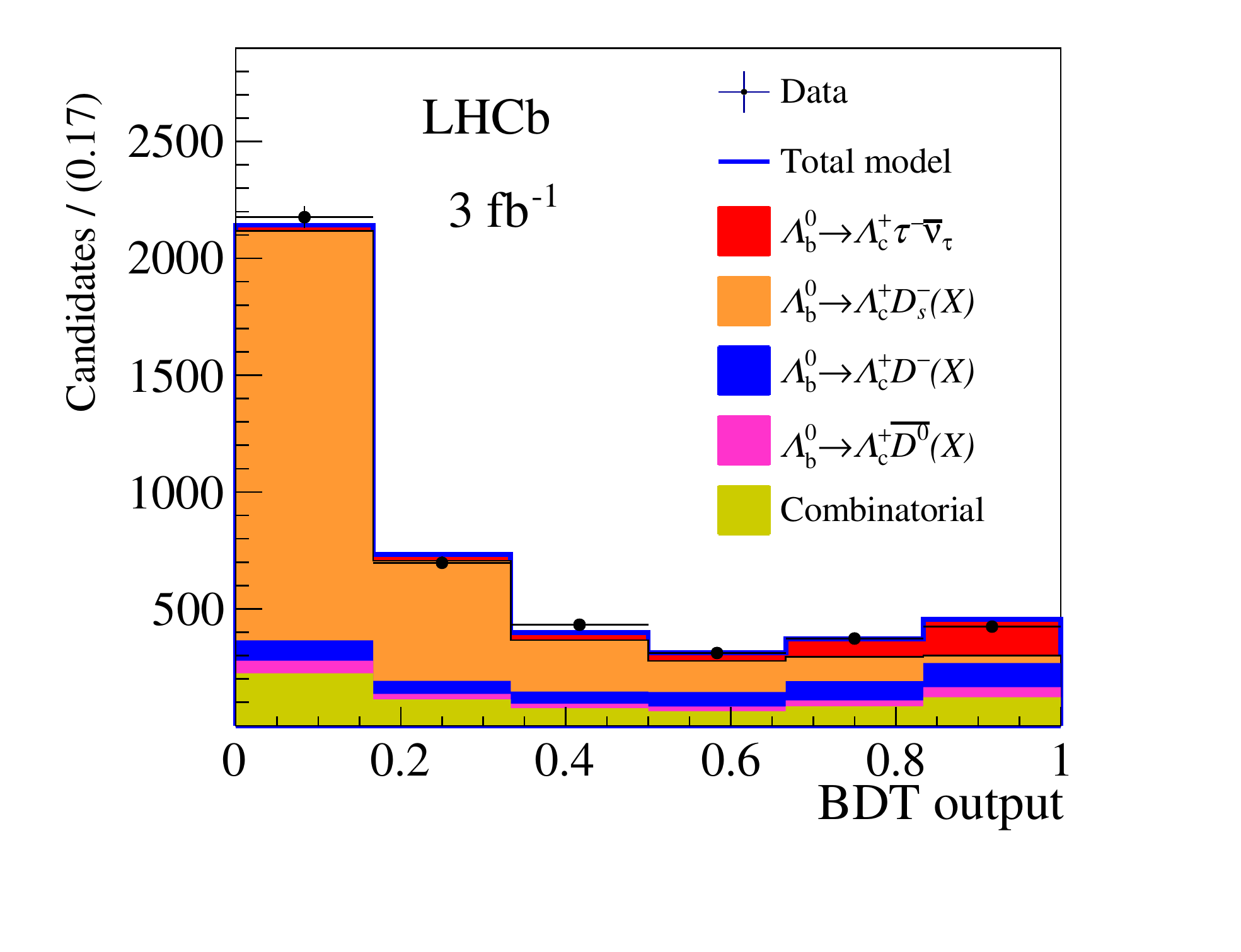}
    \end{center}
    \vspace*{-5mm}
   \caption{
     \small
 Distributions of (left) \taum decay time   and  (right) BDT output  for  \Lb\to\Lc\taum\neutb candidates.  Projections of the three-dimensional fit results are overlaid. The various fit components are described in the legend. 
  }
   \label{fig:fit_results}
 \end{figure}
\begin{figure}[!htbp]
   \begin{center}
 \includegraphics[width=0.49\textwidth]{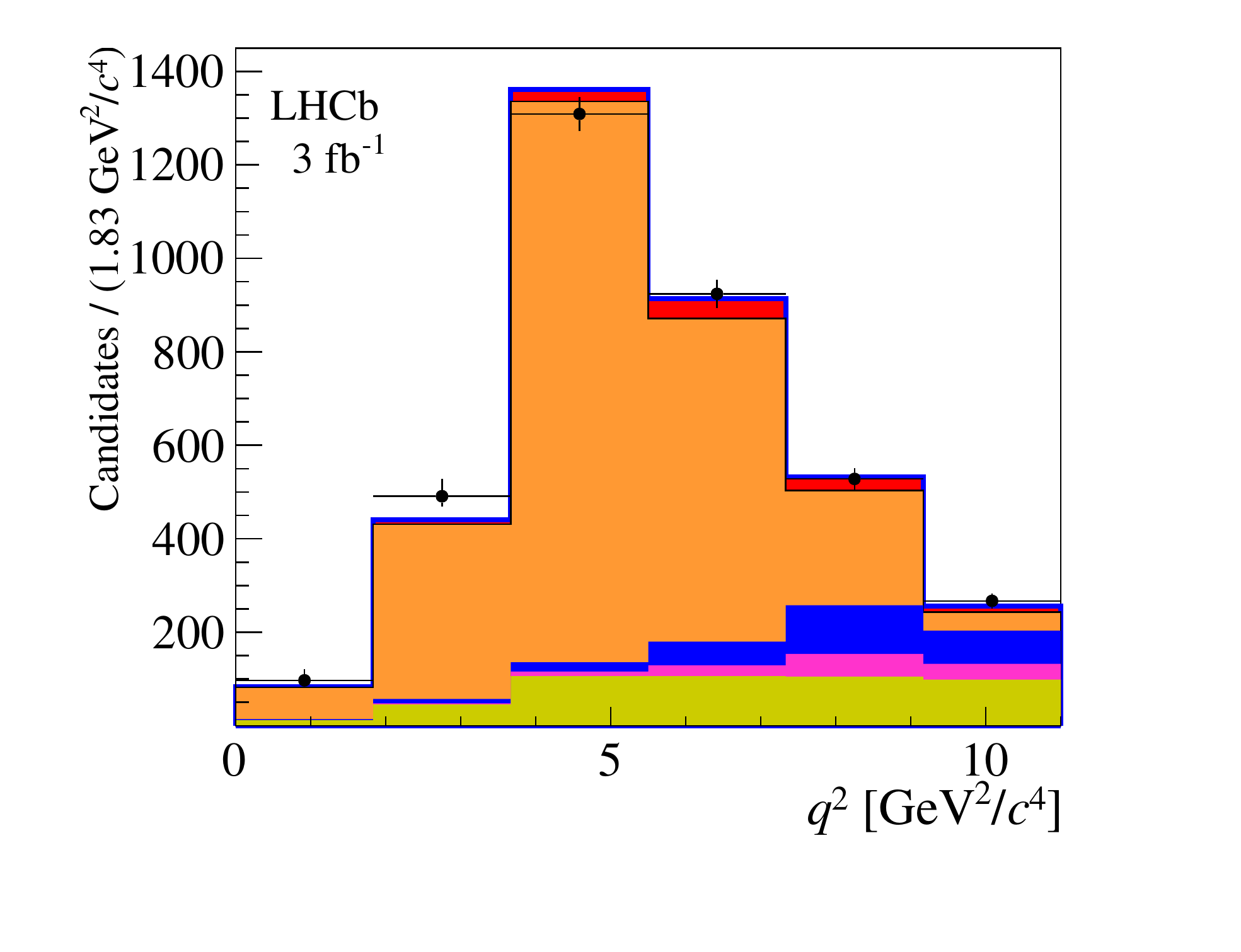}
 \includegraphics[width=0.49\textwidth]{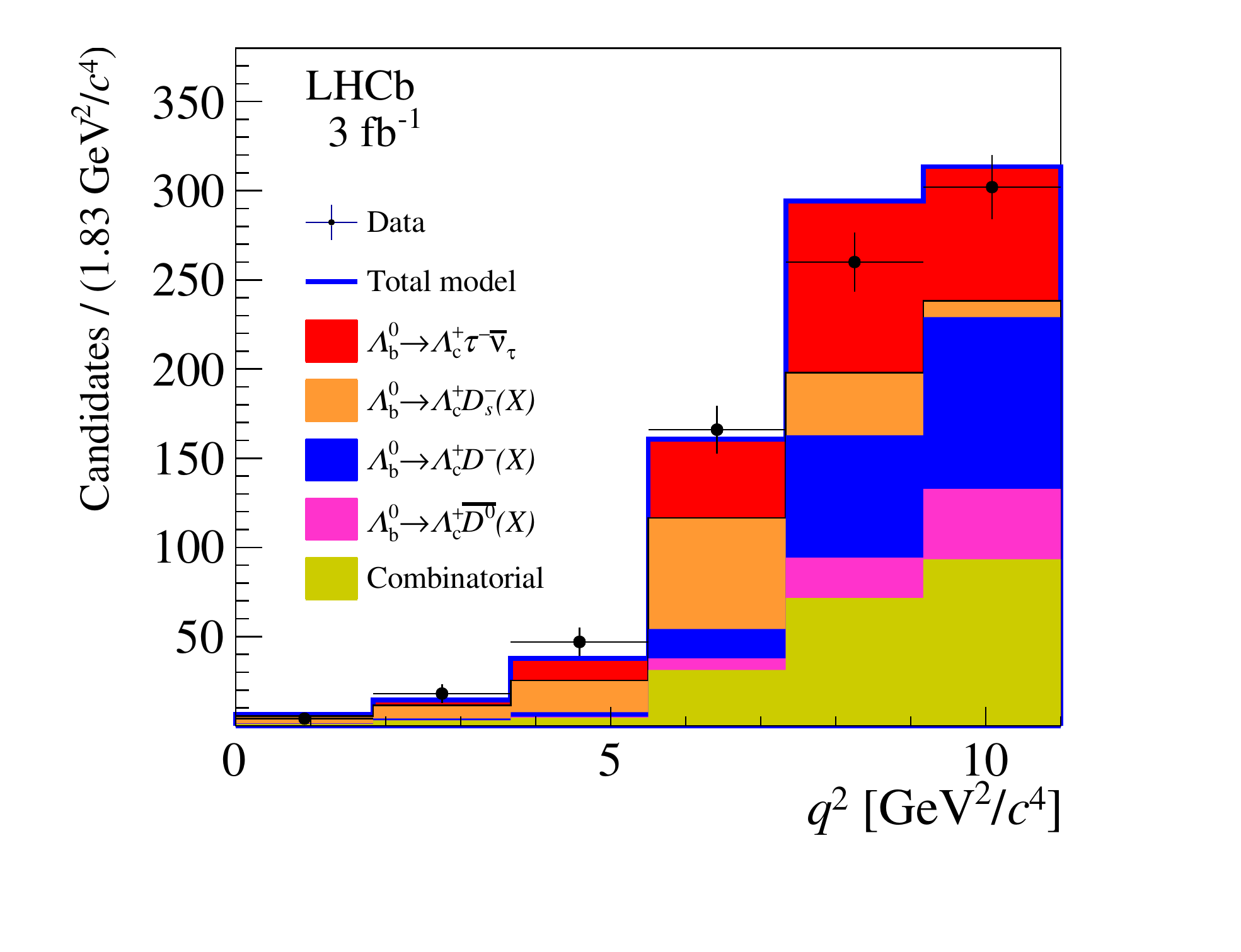}
   \end{center}
    \vspace*{-5mm}
   \caption{
     \small 
 Distributions of \qsq for \Lb\to\Lc\taum\neutb candidates having a BDT output value (left) below  and (right) above   0.66. Projections of the three-dimensional fit  are overlaid. The various fit components are described in the legend. 
  }
   \label{fig:fitq2_results}
 \end{figure}
\begin{figure}[]
    \centering
        \includegraphics[width=0.7\textwidth]{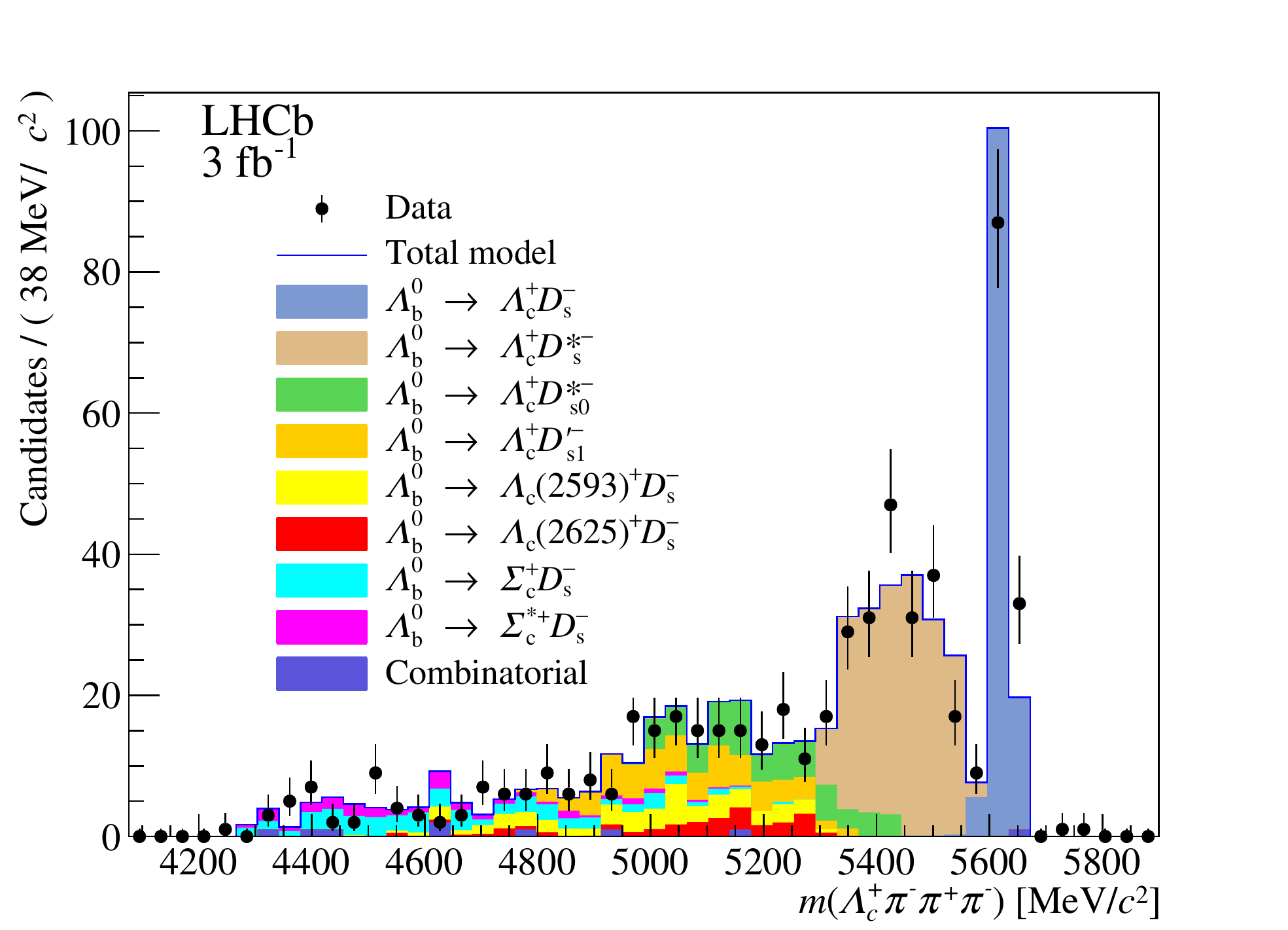}
         \vspace*{5mm}
        \caption{
    \small 
Distribution of the \Lc\pim\pip\pim invariant mass for the 
$\Lb\to\Lc\Dsm (X)$ control sample,
with $\Dsm\to\pim\pip\pim$.
The components contributing to the fit model are indicated in the 
legend.     }
  \label{fig:Ds_control_Fit}
\end{figure}

The signal yield is measured using a three-dimensional binned maximum-likelihood fit
to  $t_{\tauon}$, the BDT output, and \qsq, which are shown in Fig.~\ref{fig:fit_results} and ~\ref{fig:fitq2_results}. The fit model includes a signal component; background components due to $\B\to\Lc\Dsm (X)$,
$\Lb\to\Lc\Dm (X)$ and $\Lb\to\Lc\Dzb (X)$ decays; background due to misreconstructed \Lc candidates; and combinatorial background. Template distributions for  signal and background are obtained from simulation, with the exceptions of random $p$\Km\pip combinations and the combinatorial background, which are constructed from data-based control samples.
The signal template accounts for both $\taum\to 3\pi\nu_{\tau}$ and
$\taum\to 3\pi\piz\nu_{\tau}$ decays, where the  fraction of the former is fixed  to 78\% according to the branching fractions and selection efficiencies. 
A contribution from $\Lb\to\Lcstar\taum\overline{\nu}_{\tau}$ decays, where \Lcstar denotes  any excited charmed baryon state decaying into final states involving \Lc baryon, constitutes a feeddown to the signal. Its yield fraction is constrained to be  (10 $\pm$ 5)\% of the signal yield, derived from the \Lcstar relative abundance as measured in the $\Lb\to\Lcstar\pim\pip\pim$ decays, their respective branching fractions in the \Lc\piz\piz and \Lc\pip\pim modes~\cite{PDG2020}, and the corresponding selection efficiency obtained from simulation.
The background originating from decays of \Lb\to$\Lc\Dsm(X)$ is divided
into contributions from $\Lb\to\Lc\Dsm$,
$\Lb\to\Lc\Dssm$, $\Lb\to\Lc D^{*}_{s0}(2317)^-$, $\Lb\to\Lc D^{}_{s1}(2460)^-$, and $\Lb\to \Lcstar\Dsm (X)$ decays. 
A control sample of $\Lb\to\Lc3\pion$ candidates, where the 3\pion invariant mass is selected within 45\xspace\mevcc of  the known \Dsm mass~\cite{PDG2020} is shown in  Fig.~\ref{fig:Ds_control_Fit}. The relative yield of each of above mentioned background processes is constrained using the results of a fit to the \Lc\pim\pip\pim mass distribution.

The \Dsm decay model used in the simulation does not accurately describe the
data because of the limited knowledge of the \Dsm decay amplitudes
to 3$\pi Y$ final states. A correcting factor, taken from high precision $\Bzb\to\Dstarp\Dsm$ sample~\cite{LHCb-PAPER-2017-027}, is applied to each \Dsm branching fraction to match the \pim\pip\pim Dalitz distributions from simulation to those observed in data. 

The background originating from $\Lb\to\Lc\Dzb (X)$ decays is subdivided into two contributions,
depending on whether the 3\pion system originates from the \Dzb vertex, or
whether one pion originates from the \Dzb vertex and the other
two from elsewhere. The former contribution is constrained by 
the yield obtained from the $\Lb\to\Lc\Dzb (X)$ control
sample. The template associated to $\Lc\Dzb(X)$ background is also
validated using the data-driven sample where the \Dzb\to$ K^+3\pion$ decay is fully reconstructed. The yield of the other
$\Lb\to\Lc\Dzb (X)$ background component is a free
parameter in the fit. The yield of the $\Lb\to\Lc\Dm (X)$ background is also a free parameter and its template is validated using the data-driven sample with the \Dm meson fully reconstructed in the $\Kp\pim\pim$ mode. 

The combinatorial background is divided into two contributions, depending on whether the \Lb candidate contains a true \Lc baryon or a random $pK^{-}\pi^{+}$ combination. In the first case, the
\Lc and the 3\pion system  originate from different $b$-hadron decays.   The data sample
of wrong-sign \Lb candidates where the \Lc and the 3\pion system have the same electric
charge is used to obtain a background template. Its yield is obtained by  normalising to the right-sign data in the region where
the reconstructed $\Lc3\pion$ mass is significantly larger than the known \Lb mass~\cite{LHCb-PAPER-2017-027}. The background not
including a true \Lc baryon is parameterised  using a specific data sample originating from \Lb candidates where the \Lc candidate has a mass outside a window of 15\mevcc around the known \Lc mass~\cite{PDG2020}. 

The projections of the fit on t$_{\tauon}$ and the BDT output are shown in Fig.~\ref{fig:fit_results}. The projections on \qsq in two different BDT output ranges are shown in Fig.~\ref{fig:fitq2_results}. The signal 
yield is  $N_{\mathrm{sig}}=349 \pm 40$. The fit is repeated with all nuisance parameters related to the template shapes varying freely, while the signal yield is fixed at zero. The \chisq variation derived from the change of the fit maximum likelihood corresponds to an increase of 6.1\,$\sigma$ with respect to the default fit with freely varying signal yield. This measurement signifies the first observation of the decay  \mbox{\Lb\to\Lc$\tau^-\overline{\nu}_{\tau}$}. 
A clear separation  between signal and the main background originating from $\Lb\to\Lc\Ds(X)$ decays is obtained, as demonstrated in the BDT distribution of Fig.~\ref{fig:fit_results}.
Figure~\ref{fig:fitq2_results} shows that the $\Lb\to\Lc\Ds(X)$ background is dominant at low BDT values, while a good signal-to-background ratio is observed at high BDT output. Fig.~\ref{figsupp:bdt_taudecay}  of Supplemental Material~\cite{suppltau} shows similarly the \tauon decay time distribution for the same BDT intervals.

In order to
reduce experimental systematic uncertainties, the $\Lb\to\Lc3\pion$ decay is chosen as a normalization
channel. This 
leads to a measurement of the ratio  
\begin{equation}
\small
{\cal{K}}(\Lc)\equiv\frac{{\cal{B}}(\Lb \to \Lc \tau^-
    \neutb)
}{{\cal{B}}(\Lb \to
    \Lc3\pion)} =
    \frac{N_{\mathrm{sig}}}{N_{\mathrm{norm}}}\frac{\varepsilon_{\mathrm{norm}}}{\varepsilon_{\mathrm{sig}}}\frac{1}{{\cal{B}}(\tau^-\to
    3\pion(\piz)\neut)},
\label{eqn:kappa}
\end{equation}
where $N_{\mathrm{sig}}$ ($N_{\mathrm{norm}}$) and $\varepsilon_{\mathrm{sig}}$
($\varepsilon_{\mathrm{norm}}$) are the yield and selection efficiency
for the signal (normalization) channel, respectively. The normalization channel selection is identical to that of the signal channel, except the requirement that the 3\pion system has a larger flight distance than that of the \Lc candidate, which is not imposed. 
 The yield of the normalization mode is determined by fitting the 
invariant-mass distribution of the $\Lc3\pion$ candidates  around the
known \Lb mass~\cite{PDG2020}, as shown in Fig.~\ref{figsupp:normalization} of Supplemental Material~\cite{suppl}. A  significant contribution  from excited baryons which decay to \Lc\pip\pim, \Lc\pip, or \Lc\pim  is explicitely vetoed from the normalization channel. As a result, the 3\pion dynamics ressembles that of the signal, leading to a reduced systematic uncertainty.

A normalization yield of $N_{\mathrm{norm}} = 8584~\pm$ 102  is found, after subtraction of
a small contribution of $168\pm 20$ $\Lb\to\Lc\Dsm (\to 3\pi)$
decays. This component is estimated by fitting the 3\pion mass
distribution in the \Dsm mass region for candidates with a reconstructed \Lc3\pion mass in a window around the
known \Lb mass~\cite{PDG2020}. The normalization sample is also used to correct for differences in the \Lb production kinematics between data and simulation.
The reconstruction efficiencies for the \taum\to3\pion\neut,  \taum\to3\pion\piz\neut signal modes and normalization channel are determined using the simulation and found to be $(1.37\pm0.04)\times10^{-5}$, $(0.82\pm0.05)\times10^{-5}$, and $(11.21\pm0.11)\times10^{-5}$, respectively. The ratio of branching fractions is derived from Eq.~\ref{eqn:kappa} as 
\begin{equation*}
{\cal{K}}(\Lc) =2.46 \pm 0.27\pm 0.40,
\end{equation*}
\noindent where the first uncertainty is statistical and the second systematic.

Using ${\cal{B}}(\Lb\to\Lc 3\pi)= (6.14 \pm 0.94) \times 10^{-3}$~\cite{PDG2020} corresponding to an average of  measurements by the CDF~\cite{cdf}, and LHCb~\cite{LHCB-PAPER-2011-016} experiments, the signal branching fraction is determined as
\begin{equation*}
{\cal{B}}(\Lb\to\Lc\taum\overline{\nu}_{\tau})= {\ensuremath  (1.50 \pm 0.16 \pm 0.25 \pm 0.23)}\%, 
\end{equation*}
\noindent where the first uncertainty is statistical, the second systematic and the third is due to the  external branching fraction measurement.  
The branching fraction  ${\cal{B}}(\Lb\to\Lc\mu^-\overline{\nu}_{\mu}) = (6.2 \pm 1.4)\%$ from the DELPHI experiment~\cite{delphi} updated in Ref.~\cite{PDG2020} is used to obtain the ratio of semileptonic branching fractions $\mathcal{R}$(\Lc) as 
\begin{equation*}
{\cal{R}}(\Lc) ={\ensuremath  0.242 \pm 0.026\pm 0.040 \pm 0.059},
\end{equation*}
\noindent where the first uncertainty is statistical, the second systematic and the third is due to the  external branching fractions measurements. 
 The measured value of ${\cal{R}}(\Lc)$ is lower than but in agreement with the Standard Model prediction of 0.324~$\pm$~0.004~\cite{Bernlochner}. 

The  sources of systematic uncertainty of  ${\cal{K}}(\Lc)$  are reported in Table~\ref{tab:systematics}. For ${\cal{B}}(\Lb\to\Lc\taum\overline{\nu}_{\tau})$ and ${\cal{R}}(\Lc)$, the systematic uncertainties related to the external branching fractions are added in quadrature.
\begin{table}
  \centering
  \caption{Relative systematic uncertainties in ${\cal{K}}(\Lc)$.}
  \label{tab:systematics}
    \begin{tabular}{l c}
      \hline
      Source                          & $\delta \mathcal{K}(\Lc) /
                                        \mathcal{K}(\Lc) [\%]$ \\
      \hline

      Simulated sample size     & \hphantom{0}3.8 \\ 
      Fit bias                            & \hphantom{0}3.9 \\
      Signal  modelling               &  \hphantom{0}2.0 \\ 
      \Lb\to$\Lcstar\taum\overline{\nu}_{\tau}$ feeddown & \hphantom{0}2.5 \\ 
      $D_s^- \to 3\pi Y$ decay model         & \hphantom{0}2.5 \\
      $\Lb\to  \Lc D_s^- X$, $\Lb\to \Lc D^- X$, $\Lb\to\Lc \Dzb X$ background & \hphantom{0}4.7 \\ 
      Combinatorial background              & \hphantom{0}0.5 \\
      Particle identification and trigger corrections&\hphantom{0}1.5\\
      Isolation BDT classifier and vertex selection requirements&\hphantom{0}4.5\\
      \Dsm, \Dm, \Dzb template shapes &13.0\\
      Efficiency ratio                  & \hphantom{0}2.8\\\hline

 normalization channel efficiency (modelling of $\Lb\to\Lc 3\pi$) &\hphantom{0}3.0 \\ 
        \hline
      Total uncertainty                  & $16.5$ \\
      \hline
 
    \end{tabular}

\end{table}
The uncertainty  
due to the limited size of the simulated
samples is computed by repeatedly sampling each template with a bootstrap
procedure, performing the fit, and taking the standard deviation of the
resulting spread of $N_{\rm sig}$ values. 
 The limited size of the simulated samples also contributes
to the systematic uncertainty 
in the efficiencies for signal and
normalization modes. The systematic uncertainty associated with the signal decay model
originates from the limited knowledge of the form
factors and the $\taum$ polarization. The form factor distributions are varied in their range allowed by measurements from $\Lb\to\Lc\mun\neumb$ decays. The contribution from the relative branching fractions and
selection efficiencies of $\tau^-\to 3\pi\piz\nu_{\tau}$ and 
$\tau^-\to 3\pi\nu_{\tau}$ decays are computed by varying their ratio within their uncertainties. Potential contribution from other \taum decay modes is investigated through a dedicated simulation including all known \taum decay modes. Feed-down contribution where the \taum is produced in association with an excited charmed baryon
is computed varying in the fit the relative amount of such decays in their allowed range of ${\ensuremath (10 \pm 5)\%}$.
The uncertainty due to the knowledge of the \Dsm decay model is 
estimated by repeatedly varying the correction factors of the
templates within their uncertainties, as determined from the associated
control sample, and performing the fit. 
The spread of the fit results 
is assigned as the
corresponding systematic uncertainty.

The template shapes of the $\Lb\to\Lc\Dsm(X)$, $\Lb\to\Lc\Dzb X$ and
$\Lb\to\Lc\Dm(X)$ background modes depend on the dynamics of the
corresponding decays. A range of template deformations~\cite{LHCb-PAPER-2017-027} is performed, and the spread of the fit results is 
taken as a systematic uncertainty. The resulting uncertainty of 13\% represents the largest single source. A similar procedure is applied to the template for the
combinatorial background.
The contribution from a potential bias in the fit is explored by fitting pseudoexperiments where the signal strength is varied from its SM value to a negligible amount. 
Other sources of systematic uncertainty arise from
the inaccuracy on the yields of the various background
contributions, and from the limited knowledge of
the normalization channel modelling. The contribution from the removal of \Lcstar modes from the normalization channel is taken into account by varying the branching fractions of the various excited baryons decays within their measured range.

Systematic effects in the efficiencies for signal and normalization channels partially cancel in the ratio, with the remaining uncertainty being mostly due to the limited size of simulated sample. 
The trigger efficiency depends on the distributions of the decay time  of the \taum candidates and 
the invariant mass of the $\Lc 3\pion$ system. These distributions
differ between the signal and normalization modes, and the corresponding difference of the trigger efficiencies  is taken into account.

In conclusion, the first observation of the semileptonic decay \Lb\to\Lc\taum$\overline{\nu}_{\tau}$ is reported with a significance of 6.1$\,\sigma$, using a data
    sample of $pp$ collisions, corresponding to 3\invfb  of integrated luminosity, collected by
    the LHCb experiment. The measurement exploits  the
three-prong hadronic \taum decays with the
 technique pioneered by the LHCb experiment for the $\mathcal{R}$(\Dstarp) measurement~\cite{LHCb-PAPER-2017-027}. The ratio ${\cal{K}}={\cal{B}}(\Lb \to \Lc \taum
    \overline{\nu}_{\tau})/{\cal{B}}(\Lb \to \Lc \pi^-\pi^+\pi^-)$ is measured to be  ${\ensuremath 2.46 \pm 0.27\pm 0.40}$, where the first uncertainty is statistical and the second systematic. The branching fraction ${\cal{B}}(\Lb\to\Lc\taum\overline{\nu}_{\tau})$ is measured to be ${\ensuremath  (1.50 \pm 0.16 \pm 0.25 \pm 0.23)}\%$, where the third uncertainty is due to  external branching fraction measurements. A measurement of ${\cal{R}}(\Lc) ={\ensuremath  0.242 \pm 0.026\pm 0.040 \pm 0.059}$ is reported. 
 The ${\cal{R}}$(\Lc) ratio is found to be  in agreement with  the SM prediction. This measurement provides constraints on   new physics models, such as some of those described in Ref.\cite{Datta}, for which  large values of ${\cal{R}}$(\Lc) are allowed by existing ${\cal{R}}(D^{(*)})$ measurements.


\section*{Acknowledgements}
%
%
\noindent We express our gratitude to our colleagues in the CERN
accelerator departments for the excellent performance of the LHC. We
thank the technical and administrative staff at the LHCb
institutes.
We acknowledge support from CERN and from the national agencies:
CAPES, CNPq, FAPERJ and FINEP (Brazil); 
MOST and NSFC (China); 
CNRS/IN2P3 (France); 
BMBF, DFG and MPG (Germany); 
INFN (Italy); 
NWO (Netherlands); 
MNiSW and NCN (Poland); 
MEN/IFA (Romania); 
MSHE (Russia); 
MICINN (Spain); 
SNSF and SER (Switzerland); 
NASU (Ukraine); 
STFC (United Kingdom); 
DOE NP and NSF (USA).
We acknowledge the computing resources that are provided by CERN, IN2P3
(France), KIT and DESY (Germany), INFN (Italy), SURF (Netherlands),
PIC (Spain), GridPP (United Kingdom), RRCKI and Yandex
LLC (Russia), CSCS (Switzerland), IFIN-HH (Romania), CBPF (Brazil),
PL-GRID (Poland) and NERSC (USA).
We are indebted to the communities behind the multiple open-source
software packages on which we depend.
Individual groups or members have received support from
ARC and ARDC (Australia);
AvH Foundation (Germany);
EPLANET, Marie Sk\l{}odowska-Curie Actions and ERC (European Union);
A*MIDEX, ANR, IPhU and Labex P2IO, and R\'{e}gion Auvergne-Rh\^{o}ne-Alpes (France);
Key Research Program of Frontier Sciences of CAS, CAS PIFI, CAS CCEPP, 
Fundamental Research Funds for the Central Universities, 
and Sci. \& Tech. Program of Guangzhou (China);
RFBR, RSF and Yandex LLC (Russia);
GVA, XuntaGal and GENCAT (Spain);
the Leverhulme Trust, the Royal Society
 and UKRI (United Kingdom).

\newpage
\section*{Supplemental material  for the paper "Observation of the decay \Lb\to\Lc\taum\neutb"}
\begin{figure}[ht!]
    \centering
       \includegraphics[width=0.49\textwidth]{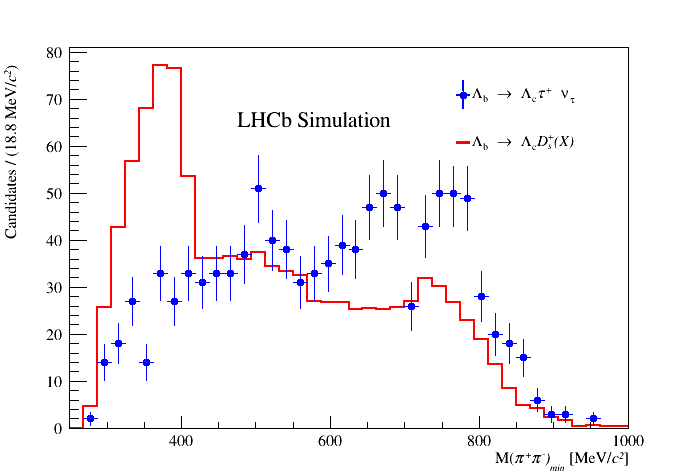} 
       \includegraphics[width=0.49\textwidth]{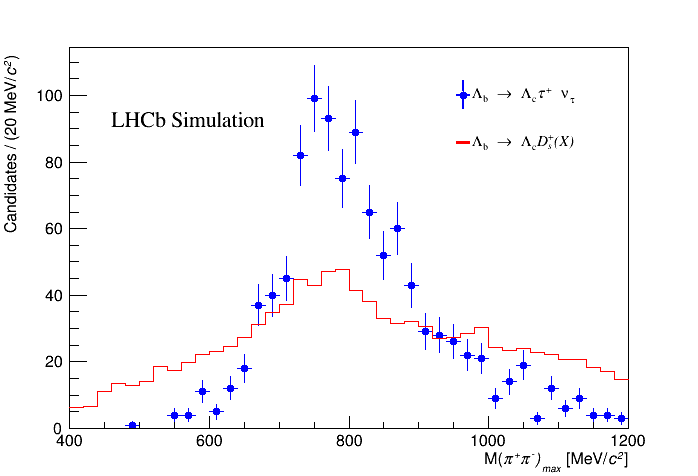}
       \includegraphics[width=0.49\textwidth]{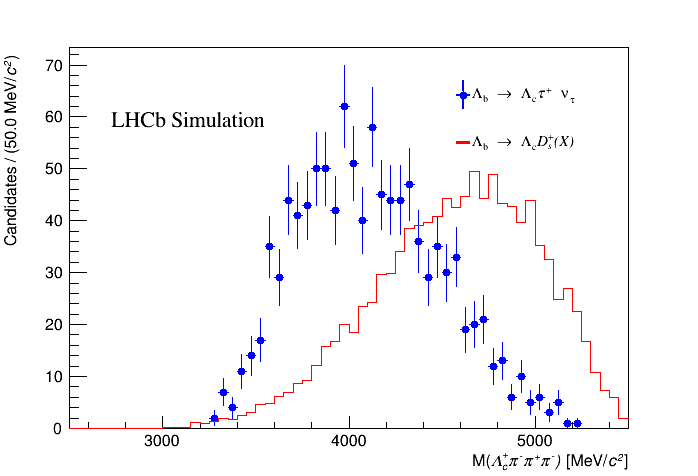}
        \caption{
    \small 
Distribution of  the (top left) minimum \pip\pim mass combination formed from the pion triplet, (top right) maximum  \pip\pim mass combination and (bottom) \Lc\pim\pip\pim mass, for simulated samples of \Lb\to\Lc\taum\neutb (blue points) and \Lb\to\Lc\Dsm(X) (red line) decays. }
  \label{figsupp:bdt_input}
\end{figure}
Fig. 4 shows the distribution of the three most significant input variables to the BDT classifier : minimum and maximum mass of the two $\pip\pim$ combinations formed from the pion triplet and $\Lc\pim\pip\pim$ mass.
Fig. 5  shows the $\tauon$ decay time  distribution for BDT output values below and above 0.66.
\begin{figure}[h!]
    \centering
       \includegraphics[width=0.49\textwidth]{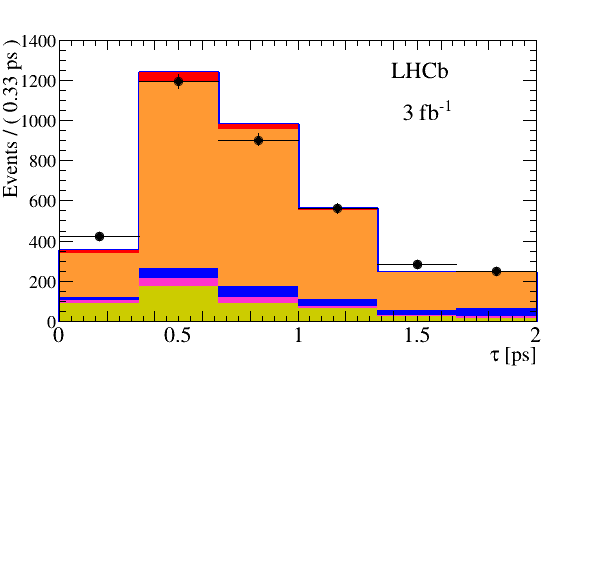} 
       \includegraphics[width=0.49\textwidth]{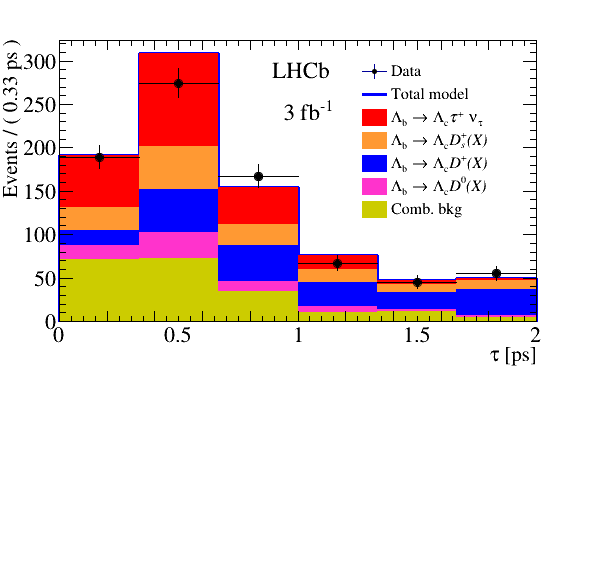}
        \caption{
    \small 
Distribution of the $\tauon$ decay time for  \Lb\to\Lc\taum\neutb candidates with (top) BDT output value below 0.66 (bottom) BDT output value above 0.66.   The various fit components are described in the legend. }
  \label{figsupp:bdt_taudecay}
\end{figure}
Fig. 6  shows the invariant-mass distribution of selected $\Lc3\pion$ candidates with a fit overlaid to extract the yield of  the normalization mode.
\begin{figure}[h!]
    \centering
        \includegraphics[width=0.8\textwidth]{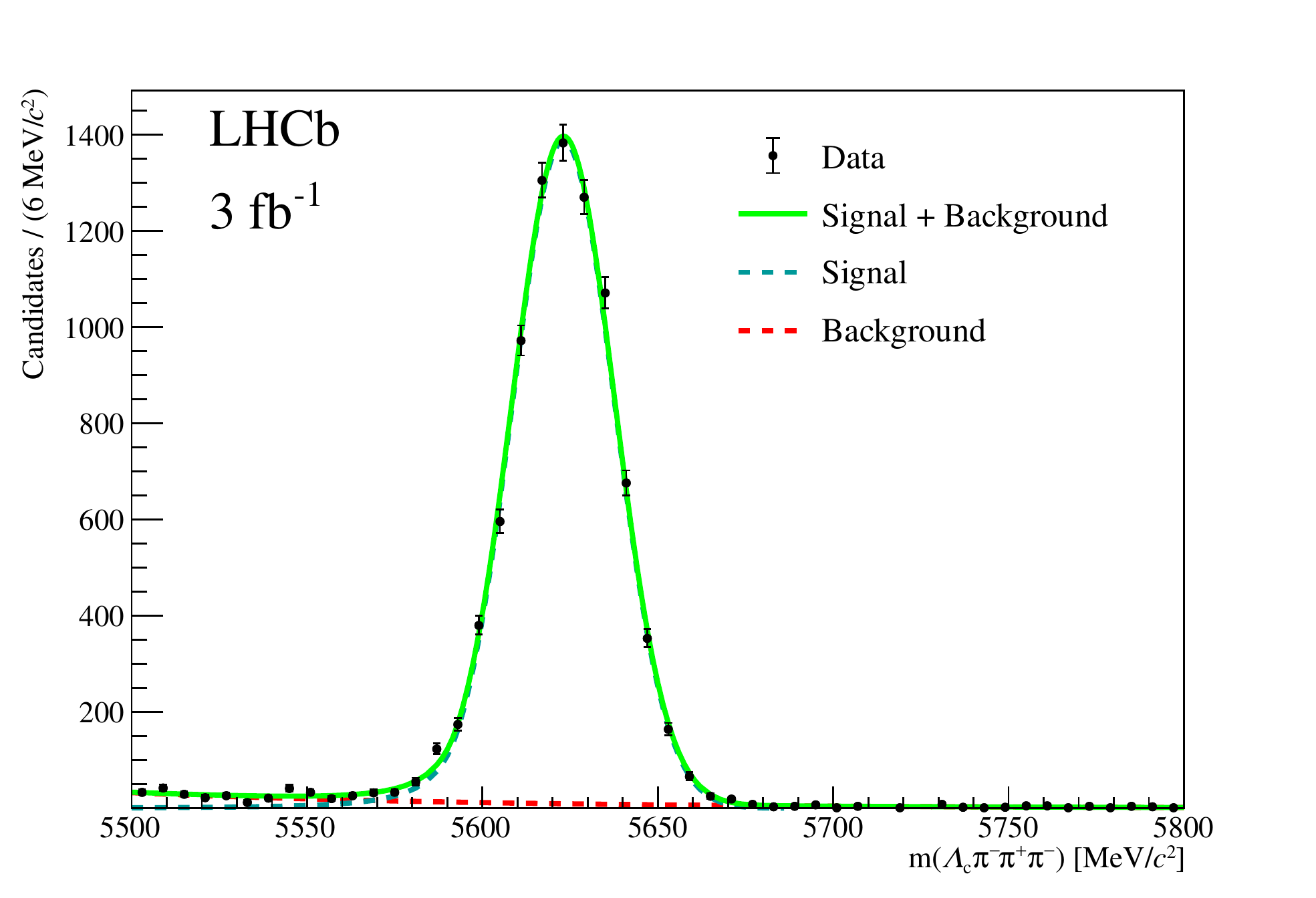}
        \caption{
    \small 
Distribution of the \Lc\pim\pip\pim invariant mass for all candidates in the normalization channel, after removal of the \Lcstar contributions. The fit components are indicated in the legend. The signal is described by a Crystal Ball (CB) function, and the background by an exponential term.}
  \label{figsupp:normalization}
\end{figure}

\newpage
\addcontentsline{toc}{section}{References}
\setboolean{inbibliography}{true}
\bibliographystyle{LHCb}
\bibliography{main,LHCb-PAPER,LHCb-CONF,LHCb-DP,LHCb-TDR,standard}

\newpage
\centerline
{\large\bf LHCb collaboration}
\begin
{flushleft}
\small
R.~Aaij$^{32}$,
A.S.W.~Abdelmotteleb$^{56}$,
C.~Abell{\'a}n~Beteta$^{50}$,
F.~Abudin{\'e}n$^{56}$,
T.~Ackernley$^{60}$,
B.~Adeva$^{46}$,
M.~Adinolfi$^{54}$,
H.~Afsharnia$^{9}$,
C.~Agapopoulou$^{13}$,
C.A.~Aidala$^{87}$
% ...

{\footnotesize \it

% ...
$^{9}$Universit{\'e} Clermont Auvergne, CNRS/IN2P3, LPC, Clermont-Ferrand, France\\
% ...
$^{13}$LPNHE, Sorbonne Universit{\'e}, Paris Diderot Sorbonne Paris Cit{\'e}, CNRS/IN2P3, Paris, France\\
% ...
$^{32}$Nikhef National Institute for Subatomic Physics, Amsterdam, Netherlands\\
% ...
$^{46}$Instituto Galego de F{\'\i}sica de Altas Enerx{\'\i}as (IGFAE), Universidade de Santiago de Compostela, Santiago de Compostela, Spain\\
% ...
$^{50}$Physik-Institut, Universit{\"a}t Z{\"u}rich, Z{\"u}rich, Switzerland\\
% ...
$^{54}$H.H. Wills Physics Laboratory, University of Bristol, Bristol, United Kingdom\\
% ...
$^{56}$Department of Physics, University of Warwick, Coventry, United Kingdom\\
% ...
$^{60}$Oliver Lodge Laboratory, University of Liverpool, Liverpool, United Kingdom\\
% ...
$^{68}$Syracuse University, Syracuse, NY, United States\\
% ...
$^{87}$University of Michigan, Ann Arbor, Michigan, United States, associated to $^{68}$\\
\bigskip
% ...
}
\end{flushleft}
\end{document}